\documentclass[11pt]{article}
\usepackage{rotating}
\usepackage{epsfig}
\usepackage{latexsym}
\usepackage{graphicx}
\usepackage{psfrag}
\usepackage{amsmath,amssymb}
\def\ltsim{\lower3pt\hbox{$\, \buildrel < \over \sim \, $}}  
\def\gtsim{\lower3pt\hbox{$\, \buildrel > \over \sim \, $}}  
\makeatletter
\def\section{\@startsection {section}{1}{\z@}{-3.5ex plus -1ex minus
 -.2ex}{2.3ex plus .2ex}{\large\bf}}
\def\subsection{\@startsection{subsection}{2}{\z@}{-3.25ex plus -1ex
minus -.2ex}{1.5ex plus .2ex}{\normalsize\bf}}
\makeatother

\newcommand{\captionfonts}{\small}
\makeatletter  
\long\def\@makecaption#1#2{%
  \vskip\abovecaptionskip
  \sbox\@tempboxa{{\captionfonts #1: #2}}%
  \ifdim \wd\@tempboxa >\hsize
    {\captionfonts #1: #2\par}
  \else
    \hbox to\hsize{\hfil\box\@tempboxa\hfil}%
  \fi
  \vskip\belowcaptionskip}
\makeatother   

\catcode`@=11
\def\marginnote#1{}
\newcount\hour
\newcount\minute
\newtoks\amorpm
\hour=\time\divide\hour
by60
\minute=\time{\multiply\hour by60 \global\advance\minute
by-\hour}
\edef\standardtime{{\ifnum\hour<12 \global\amorpm={am}
\else\global\amorpm={pm}\advance\hour by-12 \fi
 \ifnum\hour=0
\hour=12 \fi
 \number\hour:\ifnum\minute<10
0\fi\number\minute\the\amorpm}}
\edef\militarytime{\number\hour:\ifnum\minute<10
0\fi\number\minute}
\def\draftlabel#1{{\@bsphack\if@filesw
{\let\thepage\relax
 \xdef\@gtempa{\write\@auxout{\string
\newlabel{#1}{{\@currentlabel}{\thepage}}}}}\@gtempa
 \if@nobreak
\ifvmode\nobreak\fi\fi\fi\@esphack}
\gdef\@eqnlabel{#1}}
\def\@eqnlabel{}
\def\@vacuum{}
\def\draftmarginnote#1{\marginpar{\raggedright\scriptsize\tt#1}}
\def\draft{\oddsidemargin
0.0truein
 \def\@oddfoot{\sl preliminary draft \hfil
\rm\thepage\hfil\sl\today\quad\militarytime}
 \let\@evenfoot\@oddfoot
\overfullrule 3pt
 \let\label=\draftlabel
\let\marginnote=\draftmarginnote
\def\@eqnnum{(\theequation)\rlap{\kern\marginparsep\tt\@eqnlabel}
\global\let\@eqnlabel\@vacuum}
}
\catcode`@=12

\def\XXint#1#2#3{{\setbox0=\hbox{$#1{#2#3}{\int}$}
     \vcenter{\hbox{$#2#3$}}\kern-.5\wd0}}

\def\simlt{\stackrel{<}{{}_\sim}}

\def\bea{\begin{eqnarray}} \def\eea{\end{eqnarray}}
\def\be{\begin{equation}} \def\ee{\end{equation}} 

\newcommand{\promille}{%
  \relax\ifmmode\promillezeichen
        \else\leavevmode\(\mathsurround=0pt\promillezeichen\)\fi}
\newcommand{\promillezeichen}{%
  \kern-.05em%
  \raise.5ex\hbox{\the\scriptfont0 0}%
  \kern-.15em/\kern-.15em%
  \lower.25ex\hbox{\the\scriptfont0 00}}

\begin{document}

\thispagestyle{empty}

\begin{center}
\hfill CERN-PH-TH/2008-231\\
\hfill IFT-UAM/CSIC-08-82\\
\hfill UAB-FT-658

\begin{center}

\vspace{1.7cm}

{\LARGE\bf A Note on Unparticle Decays}

\end{center}

\vspace{1.4cm}

{\bf Antonio Delgado$^{\,a}$, Jos\'e R. Espinosa$^{\,b,c}$ , Jos\'e
Miguel No$^{\,b}$, Mariano Quir\'os$^{\,c,d}$}\\

\vspace{1.2cm}

${}^a\!\!$ {\em {Department of Physics, 225 Nieuwland Science Hall,
University of Notre Dame,\\ Notre Dame, IN 46556-5670, USA}}\\

${}^b\!\!$
{\em { IFT-UAM/CSIC, Fac. Ciencias UAM, 28049 Madrid, Spain}}\\

${}^c\!\!$
{\em { IFAE, Universitat Aut{\`o}noma de Barcelona,
08193 Bellaterra, Barcelona, Spain}}

{\em {and}}

{\em {ICREA, Instituci\`o Catalana de Recerca i Estudis Avan\c{c}ats,
Barcelona, Spain}}\\

${}^d\!\!$
{\em { Theory Division, CERN, Geneva 23 CH-1211, Switzerland}}

\end{center}

\vspace{0.8cm}

\centerline{\bf Abstract}
\vspace{2 mm}
\begin{quote}\small
The coupling of an unparticle operator ${\cal O}_U$ to Standard Model
particles opens up the possibility of unparticle decays into standard
model fields.  We study this issue by analyzing the pole structure
(and spectral function) of the unparticle propagator, corrected to
account for one-loop polarization effects from virtual SM
particles. We find that the propagator of a scalar unparticle (of
scaling dimension $1\leq d_U<2$) with a mass gap $m_g$ develops an
isolated pole, $m_p^2-i m_p \Gamma_p$, with $m_p^2\simlt m_g^2$ below
the unparticle continuum that extends above $m_g$ (showing that the
theory would be unstable without a mass gap). If that pole lies below
the threshold for decay into two standard model particles the pole
corresponds to a stable unparticle state
(and its width $\Gamma_p$ is zero). For $m_p^2$ above threshold the
width is non zero and related to the unparticle decay rate into
Standard Model particles. This picture is valid for any value of $d_U$
in the considered range.

\end{quote}

\vfill

\newpage

Unparticle physics was introduced in Ref.~\cite{Georgi} as the effective 
description of a conformal theory coupled to the Standard Model (SM).  
Unparticles have their origin in a hidden sector that flows to a strongly 
coupled conformal theory with an infrared fixed point below some energy
scale  $\Lambda_U$.
Since that theory is strongly coupled the anomalous dimensions can be 
large and (below the scale $\Lambda_U$) unparticle operators can have a 
dimension $d_U$ which differs sizably from its (integer) ultraviolet 
dimension. In this note we consider unparticles not charged under the SM 
gauge group and (in order to enhance its interactions with the Standard 
Model) with the lowest possible dimension. Therefore we will discuss 
scalar unparticles $\mathcal O_U$ with $1\leq  
d_U<2$~\cite{Fox:2007sy,Grinstein:2008qk}.

The conformal invariance of the unparticle sector is explicitly broken 
by its interactions with the Standard Model. Moreover, when the Higgs
field acquires a vacuum expectation value (VEV) this large breaking of
conformal invariance gives rise to a mass gap $m_g$ in the unparticle 
spectrum that consists of a continuum of states above $m_g$ \cite{DEQ}. 
The 
mass gap
plays a relevant role in the cosmology~\cite{McDonald:2007bt} and
phenomenology~\cite{Rizzo:2007xr}--\cite{Rajaraman:2008qt} of
unparticles and it should be taken into account when constraining the
unparticle theory from cosmological and experimental data.

In this paper we consider the issue of the stability of
unparticles coupled to the Standard Model or, in other words,
their possible decay into SM particles. (This is a controversial subject, see
~\cite{McDonald:2007bt,Strassler,deco,Rajaraman:2008bc}.) This issue 
should
have a great impact for unparticles in their influence on early 
Universe cosmology, in their capability as Dark Matter candidates 
and in their possible detection at high-energy colliders
through its production and subsequent decay into SM particles.
We will see that the possibility of decay,
along with the associated resonant structure, will depend on the
precise relationship between the mass gap $m_g$ and the SM threshold
of the channel to which the unparticle operator is coupled. In
particular we will consider the decay of unparticles into SM particles
via the Lagrangian coupling $ \mathcal L=-\kappa_U \mathcal O_{SM}
\mathcal O_U $ where $\mathcal O_{SM}$ is a SM operator which can
provide a channel for unparticle decay and $\kappa_U$ is a coupling
with dimension $4-d_U-d_{SM}$. Examples of such SM operators are
$F_{\mu\nu}^2$, $m_f\bar f f$ or $|H|^2$.

However, instead of focusing on a particular SM operator, we start by 
simply
considering a toy model with a real scalar $\varphi$, with bare mass $m_0$
and zero VEV, coupled to the unparticle scalar operator $\mathcal O_U$
with scaling dimension $d_U$ through the effective Lagrangian
\begin{equation}
\mathcal L_{eff}=\frac{1}{2}
(\partial_\mu\varphi)^2-\frac{1}{2}m_0^2\varphi^2
-\frac{1}{2}\kappa_U \varphi^2\mathcal O_U\ ,
\label{lagrangian}
\end{equation}
which should capture the main features of more realistic
channels.

The last term in the Lagrangian above induces a tadpole term for the 
unparticle operator at one-loop, which would trigger an unparticle 
VEV~\footnote{This tadpole is quadratically sensitive to UV physics so that 
one expects it to be of order $\kappa_U \Lambda_U^2/(16\pi^2)$.}. This is 
similar to what happens when the operator $\mathcal O_U$ is coupled to 
$|H|^2$ and the Higgs field $H$ acquires a VEV (although there the tadpole 
is a tree level effect). Here we see that this tadpole problem is more 
generic and would appear even without coupling the unparticles to the 
Higgs. It was shown in Ref.~\cite{DEQ} that in the presence of such 
tadpoles an infrared (IR) divergence appears, that has to be cutoff by an 
IR mass gap $m_g$. In the context of \cite{DEQ} the mass gap can be 
introduced in various different ways such that the conformal invariance is 
spontaneously broken along with the electroweak symmetry. Here we just 
assume that such a mass gap is provided by the theory. Of course the VEV 
of $\mathcal O_U$ in turn induces a one-loop correction to the mass of the 
field $\varphi$. We assume that this one-loop corrected mass squared is 
positive, $m^2>0$, so as to keep $\langle\varphi\rangle=0$. An alternative 
possibility is to impose the renormalization condition of zero 
unparticle tadpole at one loop so that  $\langle {\cal O}_U\rangle =0$. As 
we show later on, a non-zero mass gap will be necessary in any case.

In the presence of the mass gap $m_g$ the unparticle propagator reads
\cite{Georgi,Fox:2007sy}
\begin{equation}
-i P_U^{(0)}(s)= \frac{1}{D_U^{(0)}(s)}\equiv\frac{A_{d_U}}{2 \sin(\pi 
d_U)} \frac{1}{(-s+m_g^2-i\epsilon)^{2-d_U}}\ ,
\end{equation}
with
\begin{equation}
A_{d_U}=\frac{16
\pi^{5/2}}{(2\pi)^{2d_U}}\frac{\Gamma(d_U+1/2)}{\Gamma(d_U-1)\Gamma(2d_U)}\ 
,
\end{equation}
where we have explicitly introduced the mass gap $m_g$ that breaks
the conformal invariance. In fact in some scenarios this parameter can be 
related to the VEV of the Higgs field, as was shown in Ref.~\cite{DEQ}.
A spectral function analysis shows that, at this level, the unparticle 
spectrum is a continuum extending above the mass gap. More precisely, the
spectral function, defined as
\begin{equation}
\rho^{(0)}_U(s)=-\frac{1}{\pi}Im[-i P_U^{(0)}(s+i\epsilon)]\ ,
\end{equation}
is given by
\be
\rho^{(0)}_U(s)=\frac{A_{d_U}}{2\pi}(s-m_g^2)^{d_U-2}\theta(s-m_g^2) \ .
\ee

The polarization $\Sigma(s)$ induced in the unparticle propagator by
the one--loop diagram exchanging $\varphi$-fields can  be simply added by 
a Dyson resummation to give 
\begin{equation}
-i P_U^{(1)}=\frac{1}{D_U^{(1)}(s)}=\frac{1}{D_U^{(0)}(s)+\Sigma(s)}\ .
\end{equation}
The polarization $\Sigma(s)$ is given in the
$\overline{MS}$-renormalization scheme by~\cite{Kniehl:1993ay}
\begin{equation}
\Sigma(s)=\frac{\kappa_U^2}{32\pi^2}\left\{\log\left(\frac{\Lambda_U^2}{m^2}
\right)+2-
2 \lambda(s)\log\left[\frac{1+\lambda(s)}{\sqrt{\lambda^2(s)-1}}\right]
\right\}\ ,
\end{equation}
where $\lambda(s)=\sqrt{1-4m^2/s}$ and we have set the renormalization 
scale equal to the cutoff $\Lambda_U$. (For numerical work we fix  
$\Lambda_U=100\,m$).

The location of the unparticle resonances will be determined by the 
propagator poles 
$s=m_p^2-i m_p \Gamma_p$ in the complex $s$-plane (with $m_p$ the pole 
mass and $\Gamma_p$ its width). The polarization $\Sigma(s)$ has a branch 
cut that we take from the threshold at $s=4m^2$ to infinity along the 
real axis with the principal Riemann sheet corresponding to $0\le\theta
\leq 2\pi$, where $\theta$ is defined as $s-4m^2=|s-4m^2| e^{i\theta}$.
The second Riemann sheet is reached by shifting $\theta\rightarrow 
\theta+2\pi$.
It can be easily seen that a change in the Riemann sheet is equivalent to 
the replacement $\lambda(s)\to -\lambda(s)$. Then since the complete 
propagator is a function of $\lambda$
\begin{equation}
D^{(1)}(s)\equiv\mathcal D[s,\lambda(s)]
\end{equation}
the pole equations
\begin{equation}
\mathcal D[s,\epsilon_R \,\lambda(s)]=0
\label{polos}
\end{equation}
where $\epsilon_R=1(-1)$ correspond to solutions in the first (second)
Riemann sheet~\cite{Escribano:2002iv}.

\begin{figure}[thb]\begin{center}
\psfrag{d}[][b]{$d_U$}\psfrag{mpole}[][b]{$m_p$}
\psfrag{mu}[][b]{$m_g$}\psfrag{mpole2}[][b]{$m_p^2$}
\includegraphics[width=7.cm,height=5.cm]{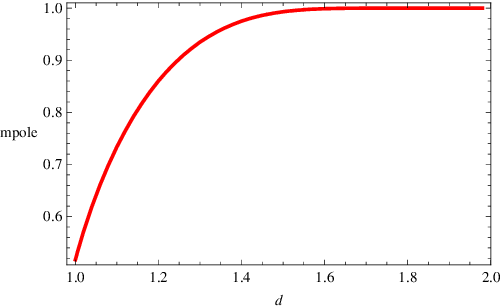}\hfill
\includegraphics[width=7.cm,height=5.cm]{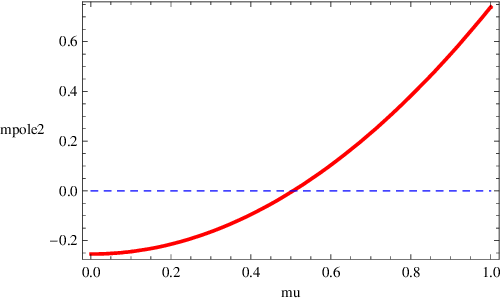}
\caption{Left panel: Plot of $m_p$ as a function of $d_U$ for
$\kappa_U=5$, $m_g=m$ and $\mu=\Lambda_U=100\, m$.  Right panel: Plot
of $m_p^2$ vs. $m_g$ for $\kappa_U=5$ and $d_U=1.2$. All
masses are in units of $m$.}
\label{mp1}\end{center}
\end{figure}

A numerical analysis of the pole equation~(\ref{polos}) shows that, 
besides the unparticle continuum, an isolated pole appears.
Note that the tree-level propagator had no pole ($m_g^2$ is not a pole but a
branch point) and therefore the pole appearance is a purely one-loop 
effect. Due to the sign of this radiative effect we find that
$m_p^2$ is always~\footnote{For $d_U$ very close to 2 one can also have 
$m_p>m_g$, but in such cases the mass difference between the pole and the 
mass gap is infinitesimal.} below $m_g^2$, but quite close to it as the 
polarization is a radiative effect: $m_p^2\simlt m_g^2$. For $m_p\leq 2m$ 
this isolated pole is real ($\Gamma_p=0$) and located in the first Riemann 
sheet. Such pole does not correspond to any decaying unparticle and it is 
entirely due to the fact that $\Sigma\neq 0$ below the threshold and could be
interpreted as an unparticle bound state. We show
in Fig.~\ref{mp1} [left panel] a plot of $m_p$ vs.~$d_U$ for $m=m_g$. In 
this plot one can see that indeed $m_p\rightarrow m_g$ for $d_U\rightarrow 
2$.

An immediate consequence of the negative mass shift responsible for  
$m_p^2<m_g^2$ is that it yields a lower bound on the scale of conformal 
breaking $m_g$. That bound is related to the masses of the Standard Model 
particles the unparticle operator is coupled to ($m$ in our case). This 
fact is 
shown by Fig.~\ref{mp1} [right panel], where the pole 
squared mass $m_p^2$ is plotted vs.~$m_g$ for $d_U=1.2$. We can see that 
the isolated unparticle pole becomes tachyonic for small values of $m_g$ 
($m_g<0.5\, m$). Moreover, this shows that in the particular limit 
$m_g\to 0$ the theory becomes unstable. Later on we give an analytical formula for
this lower bound on the mass gap.

\begin{figure}[thb]
\psfrag{d}[][b]{$d_U$}\psfrag{mpole}[][b]{$m_p$}
\psfrag{Gammapole}[][b]{$\Gamma_p$}
\includegraphics[width=7.cm,height=5cm]{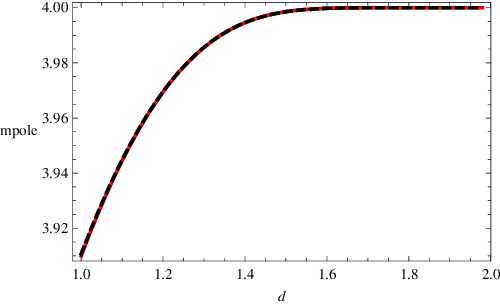}\hfill
\includegraphics[width=7.5cm,height=5.cm]{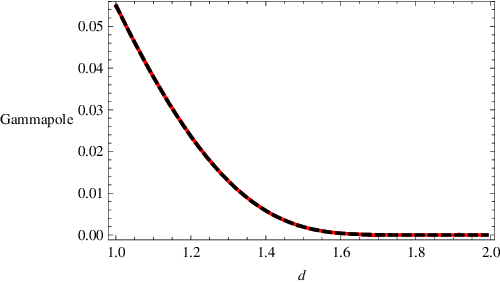}
\caption{ Left [right] panel: Plot of $m_p$ [$\Gamma_p$] as a function
of $d_U$ for $\kappa_U=5$ and $m_g=4\,m$ (thick solid). Corresponding
results based on the analytical approximation of Eq.~(\ref{anchura})
are plotted in thick dashed lines.  All masses are in units of $m$.}
\label{mp4}
\end{figure}

For $m_g> 2m$ the isolated unparticle pole is complex ($\Gamma_p>0$) and 
appears in the second Riemann sheet~\footnote{In all cases we also found 
the corresponding shadow pole~\cite{shadow} in the unphysical sheet as 
required by hermitian analyticity.}, and this now corresponds to the 
decay of a resonance. This case is exhibited in Fig.~\ref{mp4}, where 
$m_p$ and $\Gamma_p$ are plotted vs.~$d_U$ for the case $m_g=4 m$ (thick 
solid lines). Finally, since $\kappa_U^2$ is a global factor in the 
polarization, the values of $m_g^2-m_p^2$ and $\Gamma_p$ exhibit an 
approximate scaling behaviour with $\kappa_U^2$.

We want to emphasize here that there are complex pole solutions for 
all values of $d_U$ in the considered range
$1\leq d_U<2$, unlike what was claimed in Ref.~\cite{Rajaraman:2008bc}. 
In our case, nothing special happens for $d_U>3/2$ and $m_p^2$ and 
$\Gamma_p$ smoothly approach $m_g^2$ and zero respectively when $d_U\to 
2$. 

It is easy to understand analytically our results. For values of $s$ close 
to the resonance region one can approximate the complex polarization by 
the constant
\begin{equation}
\Sigma(s)\simeq
\Sigma(m_g^2)=\frac{\kappa_U^2}{32\pi^2}\left\{\log\left(
\frac{\Lambda_U^2}{m^2}
\right)+2-
\lambda(m_g^2)\left[\log\frac{1+\lambda(m_g^2)}{1-\lambda(m_g^2)}-i
\pi\right] \right\}\ ,
\end{equation}
and a simple calculation yields an analytic approximation for the pole
mass and width as
\begin{eqnarray}
m_p^2&\simeq &m_g^2- \delta m^2 \cos\alpha\ ,\nonumber\\
m_p\Gamma_p&\simeq &\lambda(m_g^2)\ \delta m^2\  |\sin\alpha |\ ,
\label{anchura}
\end{eqnarray}
where
\begin{equation}
\delta m^2\equiv \left[\frac{|\Sigma(m_g^2)|A_{d_U}}{2|\sin(\pi 
d_U)|}\right]^{\frac{1}{2-d_U}}\ ,
\end{equation}
and
\begin{equation}
 \alpha=\frac{1}{(2-d_U)}\arctan\frac{{\rm Im }[\Sigma(m_g^2)]}{{\rm Re}
 [\Sigma(m_g^2)]}\ .
\end{equation}
Figure~\ref{mp4} compares the values for $m_p$ and $\Gamma_p$ obtained
using the
analytic approximation in (\ref{anchura}) (dashed thick lines) 
with the full numerical results (thick solid lines) showing that the 
analytical approximation is excellent. We can use this approximation to 
write down analytically the lower bound on $m_g^2$ to avoid a tachyon. It 
is given by
\be
m_g^2> \left[\frac{\Sigma(0) A_{d_U}}{2|\sin(\pi 
d_U)|}\right]^{\frac{1}{2-d_U}}\ ,
\ee
with $\Sigma(0)=\kappa_U^2/(16\pi^2)\log(\Lambda_U/m)$.

We can gain further insight on the unparticle spectrum by calculating
the spectral function for the one-loop corrected propagator
\begin{equation}
\rho_U(s)=-\frac{1}{\pi}Im[-i P^{(1)}(s+i\epsilon)]\ .
\end{equation}
As we show below, this spectral function will reproduce faithfully  the
main features of the pole structure discussed previously, giving also 
information on the unparticle continuum above the mass gap.
The expression we find for this spectral function is the following:
%
\be
\rho_U(s)=\frac{1}{\pi}\frac{{\rm Im}[\Sigma(s)]}{|
D_U^{(1)}(s)|^2}+\theta(4m^2-m_p^2)\frac{\delta(s-m_p^2)}
{dD_U^{(1)}(s)/ds}
+ \theta(s-m_g^2)\frac{2 \sin^2(\pi d_U)}{\pi
A_{d_U}}\frac{(s-m_g^2)^{2-d_U}} {| D_U^{(1)}(s)|^2}\ .
\ee
%

The first term of $\rho_U(s)$ is proportional to the imaginary part of 
$\Sigma(s)$ [which contains a factor $\theta(s-4m^2)$] and thus for 
$m_p^2>4m^2$ it corresponds, through the Cutkosky rules, to a width for 
the unparticles which decay beyond the threshold. The second term (for 
$m_p^2<4m^2$) corresponds to a real pole in the first Riemann sheet, and 
should be interpreted as a stable (un)particle of mass $m_p$. Finally, the 
third term is proportional to the imaginary part of
$D_U^{(0)}(s)$~\footnote{Notice that this imaginary part is only different
from zero for $s>m_g^2$ and thus it contains a factor $\theta(s-m_g^2)$.}
and does not correspond to any unparticle decay, but gives
rise to the familiar continuous contribution to the spectral function 
above the mass gap (a similar continuum appears in the Higgs spectral 
function when the Higgs is coupled to an unparticle 
operator~\cite{Delgado:2008rq}).

When the decay $\mathcal O_U\to\varphi\varphi$ occurs it should
give rise to a resonant structure in the spectral function
$\rho_U(s)$ through the term proportional to
${\rm Im}[\Sigma(s)]$, with an approximate Breit-Wigner
distribution centered around $m_p^2$ of width $\Gamma_p$. This
should be in correspondence with the structure of the poles of the 
propagator $P^{(1)}(s)$ in the complex $s$-plane, {\it i.e.} to the 
zeroes of the function $D^{(1)}(s)$ which we have previously studied.

\begin{figure}[thb]
\psfrag{s}[][b]{$s$}\psfrag{rho}[][b]{$\rho_U$}
\psfrag{d}[][b]{$d_U$}\psfrag{K}[][b]{$K^2$}
\psfrag{Gammapole}[][b]{$\Gamma_p$}
\includegraphics[width=7.cm,height=5cm]{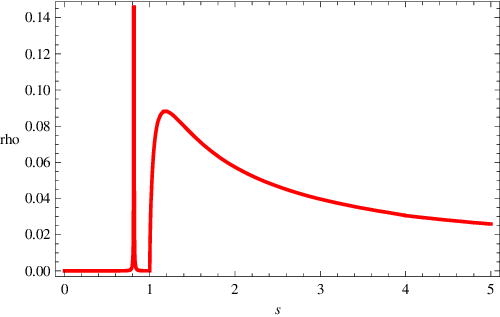}\hfill
\includegraphics[width=7.cm,height=5.cm]{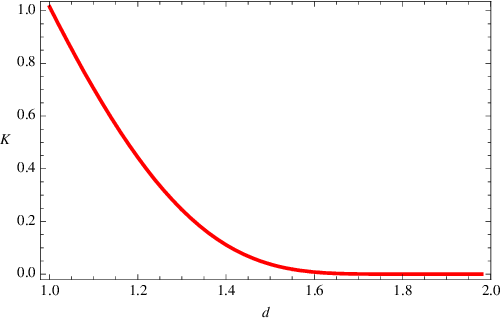}
\caption{ Left panel: Plot of $\rho_U(s)$ for $m_g=m$ , $d_U=1.25$,
$\kappa_U=5$ and $\mu=\Lambda_U=100 m$. Right panel: Plot of $K^2(d_U)$
for the same values of mass parameters. All masses are in units of $m$.}
\label{rho1}
\end{figure}

In the left panel of Fig.~\ref{rho1} we have plotted the spectral
function for the case of Fig.~\ref{mp1} in which there is no resonant
interpretation, but instead a real pole appears. We see a delta
function corresponding to that pole and a continuous component for
$s>m_g^2$. In the right panel of Fig.~\ref{rho1} we have plotted the
strength of the isolated pole, $K^2(m_p^2,d_U)$, defined as
\begin{equation}
K^2(s,d_U)=\frac{1}
{dD_U^{(1)}(s)/ds}\ .
\end{equation}

\begin{figure}[htb]
\psfrag{s}[][b]{$s$}\psfrag{rho}[][b]{$\rho_U(s)$}
\psfrag{Gammapole}[][b]{$\Gamma_p$}
\includegraphics[width=7.cm,height=5cm]{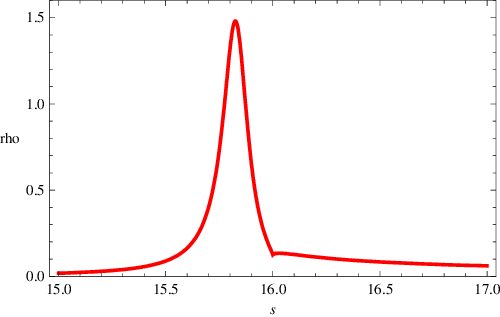}\hfill
\includegraphics[width=7.cm,height=5.cm]{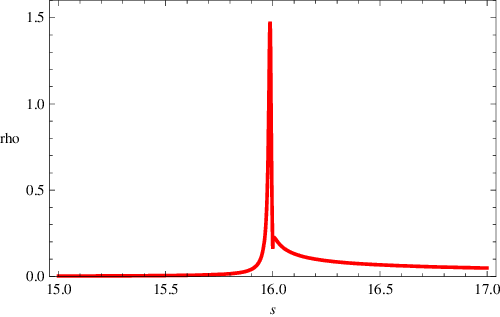}
\caption{ Left [right] panel: Plot of $\rho_U(s)$ for $d_U=1.25$
[$d_U=1.5$], $m_g=4 m$, $\kappa_U=5$ and $\mu=\Lambda_U=100 m$. All
masses are in units of $m$.}
\label{rho2}
\end{figure}

The fact that for $m_g>2m$ the pole width is sharpening for increasing
values of $d_U$ (as shown by the right plot in Fig.~\ref{mp4}) is also
shown in Fig.~\ref{rho2}, in which we plot the spectral function for
values of the dimension $d_U=1.25$ [left panel] and $d_U=1.5$ [right
panel]. In both cases we see  a clear resonant contribution
that overwhelms the continuous one. In this region and for values of
$s$ close to the value of $m_p^2$ the unparticle behaves as a
resonance

It can be easily calculated that the height of the peak is independent
of $d_U$~\footnote{This statement is true up to values of $d_U$ very
close to 2, for which the width of the resonance is zero and $m_p=m_g$
for all practical purposes.} and given by the simple expression:
\be 
\rho_U^{max}\simeq\frac{32}{\kappa_U^2\lambda(m_g^2)}\ .
\ee
Therefore, as the width goes to zero we do not recover a Dirac delta
function at $m_p$ and the resonance will be very difficult to detect
experimentally over the continuous background starting at $m_g$.

Notice that for $m_g^2>4\, m^2$ the resonant (``on-shell'') production
of unparticles would dominate the amplitude
$\varphi\varphi\to\varphi\varphi$ as it happens with ordinary exchange
of particles in the s-channel. Here the presence of unparticles should
be detected through a peak in the invariant mass distribution of the
final state similar to the case of a new particle resonance (e.g.~the
production of a $Z^\prime$).  For the case $m_p^2<4\,m^2$ the
resonance is located below the production threshold and the spectral
function is dominated by the continuous contribution, which does not
provide any decay. In that case there is no resonant production and
the production of the final state $\varphi\varphi$ will be as if
induced ``off-shell''. The presence of unparticles in the intermediate
state should be detected by the continuous enhancement of the
corresponding cross-section. This situation is reminiscent of the
familiar case of exchange of graviton Kaluza-Klein modes in ADD
theories of extra dimensions where the excess of cross section is used
to put bounds on the value of the fundamental scale.

The formalism to be used for any realistic Standard Model channel as
e.g.~$A_\mu A_\nu$, $\bar\psi_L\psi_R$ or $H^\dagger H$, is similar to
the one used in the toy model considered in this paper. In every case,
for the particular channel $\mathcal O_U\to AB$, if
$m_p^2>(m_A+m_B)^2$ the unparticle should be detected in the
corresponding cross-section through a peak in the invariant mass
distribution of the final state which should reconstruct the resonant
pole, much like the reconstruction of a $Z^\prime$ resonance. On the
contrary if $m_p^2<(m_A+m_B)^2$ then the only indirect detection of
the unparticle should be by an excess of events with respect to the
corresponding Standard Model cross-section.

In the Standard Model the only relevant (unsuppressed) operator which
can give rise to unparticle two-body decays is $\kappa_U|H|^2\mathcal
O_U$, an interaction which has been thoroughly analyzed in
Refs.~\cite{DEQ, Delgado:2008rq}. In that case the mixing in the
broken phase provided by the Lagrangian term $\kappa_U v\,h \mathcal
O_U$ gives rise to a tree-level mixing between the Higgs and
unparticles which is $\mathcal O(\kappa_U^2v^2)$. Since this mixing is
of the same order as the one loop polarization, $\mathcal
O(\kappa_U^2)$, it can be resummed in the unparticle propagator along
with $\Sigma$ leading to~\cite{DEQ} \be \mathcal D_U[s,p(s)]\to
\mathcal D_U[s,p(s)]+\kappa_U^2\,\frac{ v^2}{s-m_h^2+i\epsilon}\ , \ee
where $m_h$ is the tree-level (unmixed) Higgs mass. The analysis
should follow similar lines to those presented in this paper (after
including the extra mixing term) by just replacing $m\to m_{h}$. We
have checked that the qualitative results found in this paper do not
change after the inclusion of the Higgs-unparticle mixing.  Finally
for other channels corresponding to Standard Model particles which do
not acquire any vacuum expectation values the qualitative results
should be similar to those presented in this paper.

To summarize our results, we have studied unparticle decays into SM
particles, and showed that this possibility is controlled by the
relation between the unparticle mass gap $m_g$ and the production
threshold $m_A+m_B $ (the latter are the masses of the decay
products). When $m_g>m_A+m_B$ there is enough phase space, unparticles
can decay into SM particles, and that decay is accounted for by the
appearance of a complex pole on the unparticle one-loop resummed
propagator. If in turn $m_g<m_A+m_B$ there is still a pole but with no
imaginary part, corresponding to a stable unparticle. Finally, it
should be stressed that the pole is always below $m_g$ implying that a
theory without mass gap and coupled to a SM channel shows an
instability.  This can be interpreted as a signal that a mass gap
should be present once the unparticle is coupled to SM fields.
\section*{Appendix A. Normalization of the Spectral Function}
\setcounter{equation}{0}
\renewcommand{\theequation}{A.\arabic{equation}}

In this appendix we address the issue of the normalization of the
spectral function $\rho_U(s)$. The integral of a spectral function
along the real axis is determined by the normalization of the state
being considered. For instance, if we take the tree level spectral
function $\rho_U^{(0)}$ we would write
\be N_U^{(0)}=\int_0^\infty
\rho_U^{(0)}(s)\ ds = \langle U | U \rangle \ , 
\ee 
where $|U\rangle$ represents the unparticle operator ${\cal
O}_U$. This integral turns out to be UV divergent as corresponds to a
non-normalizable state $|U\rangle$. Strictly speaking one should
cut-off this integral at a scale of order $\Lambda_U$, beyond which
the theory leaves the conformal regime. The normalization integral
scales as
\be
N_U^{(0)}(\Lambda_U)=\int_0^{\Lambda^2_U} \rho_U^{(0)}(s)\ ds \sim
(\Lambda_U^2)^{d_U-1} \ .  
\ee 
By using the Cauchy theorem (and the absence of complex poles in the
principal Riemann sheet) one can see that $N_U^{(0)}(\Lambda_U)$ is
proportional to the integral of the propagator $P_U(p^2)$ along a
circle of radius $\Lambda_U^2$ so that its scaling is directly
dependent on the UV behaviour of such propagator.

After including one--loop polarization effects the shape of the
spectral function is affected but it keeps the same normalization as
before. In fact, defining
\be
N_U(\Lambda_U)=\int_0^{\Lambda_U^2} \rho_U(s)\ ds \ ,
\ee
one finds
\be
\frac{N_U}{N_U^{(0)}}=1+{\cal O}\left(\frac{m_g^2}{\Lambda_U^2}\right)
+{\cal O}\left(\frac{\kappa_U^2}{(\Lambda_U^2)^{2-d_U}}
\log{\Lambda_U^2}\right)\ ,
\ee
which tends to 1 for $\Lambda_U^2\gg m_g^2, (\kappa_U^2)^{1/(2-d_U)}$.

One can also show that 
\be
N_U(\Lambda_u)-N_U^{(0)}(\Lambda_U)=
\int_0^{\Lambda_U^2}[\rho_U(s)-\rho_U^{(0)}(s)]\ ds \sim 
(\Lambda_U^2)^{2d_U-3}\ ,
\ee
which tends to 0 for $d_U<3/2$, case in which the equality of the 
normalizations holds also in this stronger sense.

\section*{Acknowledgments}

\noindent
We are indebted to Rafel Escribano for useful discussions about the
pole structure of the propagator. J.M.N.~thanks the Department of
Physics of the University of Notre Dame for hospitality during the
last stage of this work. Work supported in part by the European
Commission under the European Union through the Marie Curie Research
and Training Networks ``Quest for Unification" (MRTN-CT-2004-503369)
and ``UniverseNet" (MRTN-CT-2006-035863); by the Spanish
Consolider-Ingenio 2010 Programme CPAN (CSD2007-00042); by a Comunidad
de Madrid project (P-ESP-00346); and by CICYT, Spain, under contracts
FPA 2007-60252 and FPA 2008-01430.

\end{document}